# An analytical relation between Weibull's and Basquin's laws for smooth and notched specimens and application to constant amplitude fatigue

P. D'Antuono[1,2,]

‡

## Abstract

Starting from the classical definition of stress-life Wöhler curve in the form of Basquin's law, an analytical procedure for the calibration of the four parameters Wöhler curve (Weibull's law) for a plain specimen is proposed. The obtained parameters are then adjusted by means of an additional slope factor preserving the inflection point of the curve while changing its slope in order to model the experimental observations in which an increase of the scatter in life prediction is observed when reducing the stress amplitude. The same approach has then been adopted to calibrate the Weibull's law parameters for a notched specimen, and the fitting slope factor has been found to be a value that changes with the material but remains constant with the stress concentration factor. The findings have been validated with existing experimental data on 2024-T3 aluminum alloy and normalized SAE 4130 steel.

## Nomenclature

$\hat{C}_1, \hat{C}_2$: least squares method fitting constants.

$\bar{X}, \bar{Y}$: average values of X and Y.

$\hat{Y}$: estimated value from lest squares.

$\bar{a}$: Basquin fatigue strength coefficient at 1 cycle.

a, b, B: Weibull's law parameters.

$A_N$: Neuber's constant.

$A_P$: Peterson's material constant.

$\bar{b}$: Basquin exponent.

$\check{b}, \check{a}, \check{B}, \check{\alpha}$: Corrected Weibull's law parameters.

$C_{\bar{a}}$: coefficient of variation.

$f_a$: slope factor.

K: multiplier of the standard error.

$K_f$: effective stress concentration factor.

$K_n$: technical stress concentraiton factor.

$K_t$: theoretical elastic stress concentration factor.

N: number of cycles.

$N_e$: number of cycles at $S_e$ for Basquin's law.

$n_f$: number of experimental tests.

$N_i$: inflection point in Weibull's law.

$N_u$: number of cycles at $S_u$ for Basquin's law.

q: notch sensitivity factor.

S: stress.

$S_e$: fatigue limit stress.

$S_u$: ultimate tensile strength.

X, Y: variables of the least squares regression line.

$Y_L$: lower limit of Y.

α: exponent of Weibull's law in S/log(N) coordinates.

$\Delta^2$: sum of the squared deviations.

ϵ: random error.

ρ: notch root radius.

σ: sample standard error of $Y_j$ on $X_j$.

$\sigma^2$: Standard deviation of experimental data.

---

[1] Politecnico di Bari, Department of Mechanics, Mathematics & Management, Viale Japigia, 182, Bari, Bari, IT 70126
[2] Leonardo Helicopters Division Cascina Costa, Fatigue Office, Via Giovanni Agusta, 520 Samarate, Varese, IT 21017
‡ Email at: pietro.dantuono@poliba.it

1. Introduction

The advent of the First Industrial Revolution in the 18th century caused a dramatic increase of the frequency of work of the machines; this event could be identified as the trigger for the widespread study of fatigue in the last two centuries. Indeed, even if it has been known for centuries that repeatedly applying a load causes early ruptures, as it used to happen to long distance travelling boats, it was only around 1830s that the engineers and scientists started investigating how a load much lower than the material strength can cause failure if applied many times. The first reference to the fatigue of metals dates to 1829 when Poncelet, a French military engineer, used the adjective "tired" to describe steels under stress and assumed that steel components experience a decrease of durability when they undergo repeated variable loads. In 1837 the first document in history relating to a fatigue test was published. The test was recorded by Albert [1], a German mining administrator and it was aimed to understand the causes of the failure of the conveyor chains in the Clausthal mines in 1829. In 1843 Rankine [2] and York [3], [4] focused their attention on railway axles thanks to the establishment of the Her Majesty's Railway Inspectorate instituted due to the increasing number of accidents, amongst which the so-called Versailles disaster where almost 60 people lost their lives due to the failure of the axle tree of the first engine on the 5th October 1842. Anyway, the term fatigue was coined only in 1854 by the Englishman Braithwite [5], who discussed the fatigue failures of multiple components as water pumps, brewery equipment and, of course, railway axles. Many other English and German studies on the deterioration of railway axles were made in those years [6], [7] [8], [9], but the work of Wöhler, Royal "Obermaschinenmeister" of the "Niederschlesisch-Mährische" Railways in Frankfurt an der Oder, was the milestone paving the way for the modern conception of fatigue testing and interpretation of results. In 1858 [10] and 1860 [11] August Wöhler measured for 22,000 km the service loads of railway axles with deflection gages personally developed and from his studies concluded that "If we estimate the durability of the axles to be 200,000 miles with respect to wear of the journal bearings, it is therefore only necessary that it withstands 200,000 bending cycles of the magnitude measured without failure". Such statement represents the first suggestion for a safe life design philosophy with retirement time (or distance travelled). Wöhler then calculated the stresses deriving from service loads and concluded that the higher the stress amplitude is, the more detrimental influence on the axle will be, plus a tensile mean stress anticipates the rupture. Furthermore, he even stated that a replacement of the axle would have been necessary if radial flaws propagated up to 20 mm into the material, and this procedure could be interpreted as an ancestor of the flaw tolerant safe life methodology [12]. Anyway, Wöhler's test results were tabulated and not plotted until 1875, when Spangenberg [13] adopted unusual linear axes to present these data. Furthermore, stress vs. number of cycles (S/N) curves were addressed as Wöhler curves only in 1936 by Kloth and Stroppel [14]. The idea to plot many fatigue test data, including the 60 years old Wöhler's experiments, in logarithmic axes and interpolate them with a power law is from 1910 by Basquin [15]. The equation relating maximum or alternate stress with the number of cycles is

$$S = \bar{b} \cdot N^{\bar{a}} \qquad (1)$$

Where S denotes the stress, alternate or maximum, N the number of cycles and $\bar{a}$ and $\bar{b}$ are constants depending on the material. Later such law took his name. The proof of existence of a fatigue limit was given four years later by Stromeyer [16]. In 1914 he conducted tests on small scale specimens in order to reduce to a minimum the difference in terms of chemical composition and mechanical properties in the component under test. The specimens were loaded in bending and twisting moment and the stress plotted against the empirical formula $(10^6/N)^{1/4}$ resulted in a straight line corresponding to the fatigue limit. Stromeyer also theorized that if the maximum stress would have exceeded the yielding, the constant slope ¼ would have certainly changed. The final form of Stromeyer's law simply adds a constant term to equation (1):

$$S = \bar{b} \cdot N^{\bar{a}} + S_e \qquad (2)$$

Where $S_e$ is the fatigue limit and for the same material $\bar{a}$ and $\bar{b}$ defined in equation (2) have different values from the corresponding quantities defined in equation (1). In 1924 Palmgren [17] in his studies for the ball bearings life estimation introduced a term B in equation (2), i.e.

$$S = b \cdot (N + B)^a + S_e \qquad (3)$$

The B introduces an inflection point in the equation both when plotted in S/log(N) and when plotted in log(S)/log(N) axes, consequently low cycle fatigue data are better fitted by this model. Indeed, as suggested by Shanley [18] in the 1956 "Colloquium on Fatigue" [19], Basquin's law fails to model low cycle fatigue since it does not predict correctly the strain at high stresses, while by using equation (3) one supposes that a stress close to the ultimate tensile strength causes much lower strain, hence it can be applied a certain number of times without failure. This is also confirmed by Epremian and Mehl [20] that in 1952 showed that an S/N diagram when the alternate strength is close to the ultimate tensile strength, can be fit with very good agreement by a probability scale instead of a logarithmic scale and this suggests that at high stresses alternating plastic strain dependence on stress amplitude is primarily of statistic nature. The model found by Palmgren has been widely used by Weibull since 1949 [21]. For this reason, equation (4) will be addressed here as Weibull's law. Engineer and mathematician, Ernst Hjalmar Waloddi Weibull (1887-1979) gave a huge contribution to material science and statistics in his prolific scientific career. Concerning fatigue, there are tens of documents, many of which are ICAF (International Committee on Aeronautical Fatigue and Structural Integrity) proceedings [22]–[25] and reports to the Aeronautical Research Institute of Sweden (FFA) [26]–[30]. His contribution to the field is principally, but not only, related with the statistical aspects of fatigue. As regards equation (3), Weibull in his book [31] states, in disagreement with Moore and Kommers [32], that the knee that seem to show S/N diagrams "is an accidental phenomenon caused by the joint effect of a large scatter in fatigue life and too small number of observations". Thence, for the reasons just evidenced the smooth Weibull's S/N curve model is certainly a more realistic form of modelling stress life fatigue data with respect to a Basquin's law truncated at the fatigue limit. Similarly, in crack growth analysis the NASGRO equation in its original and modified forms [33]–[35] is a much more realistic representation of the Paris' law[36] truncated below the threshold stress intensity factor and above the limit stress intensity factor. Nevertheless, Basquin's law has been extensively used by most of the researchers and engineers until these days because of its simplicity and consequently most of the fatigue databases for S/N curves available in literature are given in the form of Basquin's law truncated at the fatigue limit. To the author's knowledge, nobody before has ever tried to relate Basquin's and Weibull's laws and this shall be done in this work. The usefulness of this exercise can be found in the fact that many stress life models are based on Basquin's law, thus finding an analytical relationship with Weibull's law could make simpler to rewrite these models by means of a more sophisticated and realistic model, plus an entire database of Basquin's law coefficient might be converted into Weibull's law by simply applying an analytical formula. Moreover, as also Weibull states in his book [31], tuning the parameters of Weibull's law is not as immediate as doing the same thing with a pure power law. Indeed, Weibull suggests a graphical and a semi-anbalytical trial and error procedure to derive the parameters for his equation from experimental data. With the model described here, one could find the power law which interpolates the data in the high cycle fatigue regime and then the calculation of the four parameters is straightforward. Finally, the important matter of developing a unique S/N curve model based on Weibull's law which can account for the notch effect is addressed.

2. Wöhler curve in the form of Basquin's law

When dealing with Basquin's law, it must be considered that in its truncated form it depends on four parameters, i.e.

$$\begin{cases} S(N) = & S_u = \bar{b} \cdot N_u^{\bar{a}} & 1 \leq N \leq N_u \\ S(N) = & \bar{b} \cdot N^{\bar{a}} & N_u < N \leq N_e \\ S(N) = & S_e = \bar{b} \cdot N_e^{\bar{a}} & N > N_e \end{cases} \quad (4)$$

Here the four parameters are the slope $\bar{a}$, the fatigue strength at 1 cycle (or at one reversal) $\bar{b}$, the ultimate tensile strength $S_u(N_u)$ and the fatigue limit $S_e(N_e)$. Equation (4) corresponds to a piecewise straight line in a log-log plot. Basquin's law is also currently the most used in the field of research since because of its simplicity it can be easily manipulated. Also the author used Basquin's law in his works to derive new simplified models to predict variable amplitude fatigue life [37] or to derive another smoothed model which could account for the notch effect in the case of a plate with a circular holes [38].

## 3. Wöhler curve in the form of Weibull's law

Finding an analytical link between Basquin's and Weibull's laws does not introduce new variables and complexity to the problem since in both cases four parameters are necessary to draw the Wöhler curve; on the contrary, the user will be able to condensate a piecewise function into a single equation and consequently, no further experimental analysis should be needed to re-characterize a material by means of Weibull's S/N curve model. In this way, Weibull's law can be intended as a direct smoothing of the truncated power law. Weibull's law was defined such that three constraints had to be satisfied when plotting it in semilogarithmic coordinates: (i) slope equal to zero for N→0, (ii) slope equal to zero for N→∞ and (iii) the curve must show an inflection point. The same constraints must be valid in a log-log plot. The law along with its derivatives w.r.t. N is

$$\begin{cases} S(N) = & b \cdot (N+B)^a + S_e & (a) \\ S'(N) = & \dfrac{dS(N)}{dN} = b \cdot a \cdot (N+B)^{a-1} & (b) \\ S''(N) = & \dfrac{d^2 S(N)}{dN^2} = b \cdot a \cdot (a-1) \cdot (N+B)^{a-2} & (c) \end{cases} \quad (5)$$

The parameters $a$, $b$, $B$, $S_e$ must be calculated such that the new S/N curve respects the constraints defined by the Basquin's law. On this purpose, it is assumed that the slope of the Basquin's law matches the first derivative of Weibull's law in the inflection point $N_i$ of the log-log plot, i.e. $\log_{10} S(N)_{/\log_{10} N}\big|_{N=N_i} = \bar{a}$.

$$\begin{cases} \log_{10} S(N)_{/\log_{10} N}\big|_{N=N_i} = & \bar{a} \\ \log_{10} S(N)_{/\log_{10} N \ \log_{10} N}\big|_{N=N_i} = & 0 \end{cases} \quad (6)$$

Using the equations defined in Appendix gives

$$\begin{cases} \dfrac{b \cdot a \cdot (1 + B/N_i)^{a-1}}{b \cdot (1 + B/N_i)^a + S_e} = & \bar{a} & (a) \\ (N_i + B) \cdot (1 - a) + N_i \cdot (a - 1) = & 0 & (b) \end{cases} \quad (7)$$

The hypothesis on the slope of the inflection point is then combined with the static failure and the fatigue limit conditions, i.e.:

$$\begin{cases} b = & (S_u - S_e) \cdot B^{-a} & (a) \\ S(N \to \infty) = & S_e & (b) \end{cases} \quad (8)$$

Equation (7)(b) can be expressed as function of the auxiliary variable $\alpha = B/N_i$ corresponding to the exponent of Weibull's law in S/log(N) coordinates (it is a positive number since Weibull uses a minus sign at the exponent in his notation). With some simple passages the system of equations is then solved for $\alpha$ giving the following implicit equation

$$\left(\frac{\alpha+1}{\alpha}\right)^{\alpha+1} = -\frac{\bar{a} S_e}{S_u - S_e} \quad \text{for } 0 < \alpha < 1 \tag{9}$$

Once numerically solved, equation (9) gives the ratio $\alpha = B/N_i$. Then, the value of $N_i$ is retrieved by considering that both the classical S/N curve and the smoothed one pass by $(N_i, S(N_i))$, i.e. equation (4) has to be equal to equation (5)(a) for $N = N_i$

$$\bar{b} N_i^{\bar{a}} = b \cdot (N_i + B)^a + S_e \tag{10}$$

or

$$b = \frac{\bar{b} N_i^{\bar{a}} - S_e}{N_i^a \cdot (1 + \alpha)^a} \tag{11}$$

By using again Equation (7)(a) and the definition of $\alpha$, an explicit expression of the inflection point $N_i$ is found

$$N_i = \left(\left(\frac{1}{\alpha} + 1\right)^a \cdot \frac{(S_u - S_e)}{\bar{b}} + \frac{S_e}{\bar{b}}\right)^{1/\bar{a}} \tag{12}$$

The value of b can be found from equation (8)(a) for N=0 or from equation (11)s. From the above passages, the complete set of equations which "converts" Basquin's law into Weibull's law is

$$\begin{cases} \left(\frac{\alpha+1}{\alpha}\right)^{\alpha+1} = -\frac{\bar{a} S_e}{S_u - S_e} & \text{(a)} \\ a = \alpha \cdot (\bar{a} - 1) + \bar{a} & \text{(b)} \\ B = \alpha \cdot \left(\left(\frac{1}{\alpha} + 1\right)^a \cdot \frac{(S_u - S_e)}{\bar{b}} + \frac{S_e}{\bar{b}}\right)^{1/\bar{a}} & \text{(c)} \\ b = (S_u - S_e) \cdot B^{-a} & \text{(d)} \end{cases} \tag{13}$$

Note that for $S_e=0$ (no fatigue limit) the smoothed S/N curve shows no inflection point which implies $\alpha \to 0$, and the problem is again solvable in closed form. For a full analytical solution of the system shown in (13), an approximate form of Equation (13)(a) should be found. The basic consideration is that $\left(\frac{\alpha+1}{\alpha}\right)^{\alpha+1} = \left(\frac{\alpha+1}{\alpha}\right) \cdot \left(\frac{\alpha+1}{\alpha}\right)^\alpha$ and the approximate form to be found must be valid in the range $0<\alpha<1$. Hence, by studying the function $\left(\frac{\alpha+1}{\alpha}\right)^\alpha$, one finds that in the range of interest it can be approximated by an equilateral translated hyperbola $\frac{A_1 x + A_3}{A_2 x + A_3}$, where the coefficients $A_1=5$, $A_2=2$ and $A_3=1$ minimize the error between the correct and the approximated form, with a maximum error lower than 2.5%, as shown in Figure 1. Thus, the full analytical approximated solution to the classical Wöhler curve smoothing problem proposed in this work is

$$\begin{cases} \alpha = \dfrac{6-\gamma+\sqrt{(6-\gamma)^2+4(2\gamma-5)}}{2(2\gamma-5)} & (a) \\ a = \alpha\cdot(\bar{a}-1)+\bar{a} & (b) \\ B = \alpha\cdot\left(\left(\dfrac{1}{\alpha}+1\right)^a\cdot\dfrac{(S_u-S_e)}{\bar{b}}+\dfrac{S_e}{\bar{b}}\right)^{1/\bar{a}} & (c) \\ b = (S_u-S_e)\cdot B^{-a} & (d) \end{cases} \qquad (14)$$

Where $\gamma = -\bar{a}S_e/(S_u-S_e)$.

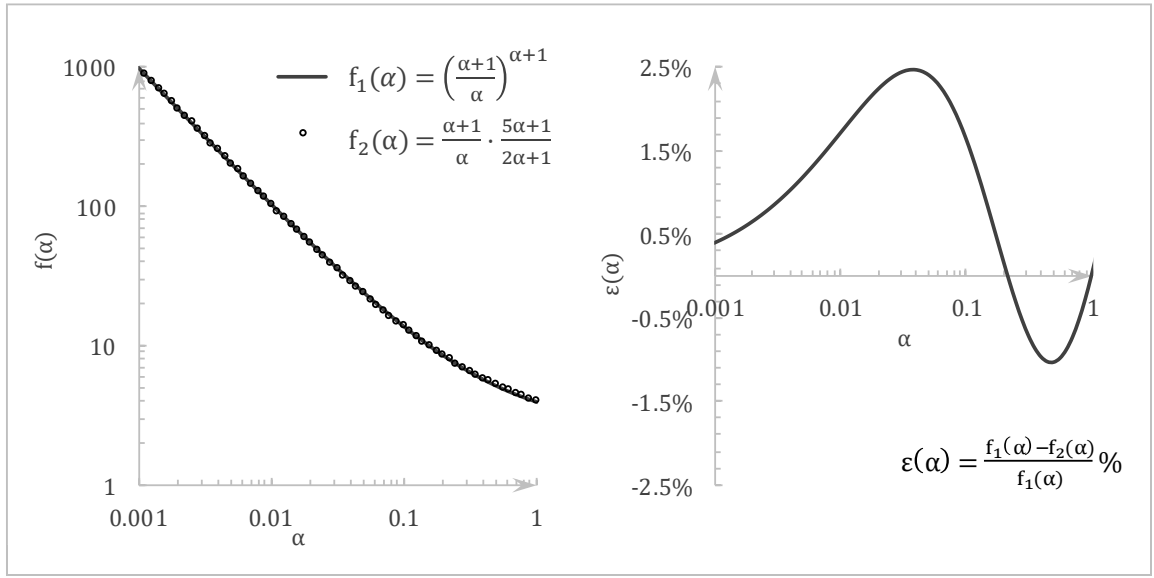

Figure 1 – Approximate form of $\left(\dfrac{\alpha+1}{\alpha}\right)^{\alpha+1}$ (left side) and relative error between $\left(\dfrac{\alpha+1}{\alpha}\right)^{\alpha+1}$ and $\dfrac{5\alpha+1}{2\alpha+1}$ (right side) in the range $0<\alpha<1$

## 4. Discussion of the results

As written in the introduction, all the stress-life material data available in literature in terms of Basquin's law can be easily recalculated by means of the smoothing technique just explained providing Weibull's law data. As an example, consider a generic steel, in this case the material data for the two forms of S/N curve are:

| | Basquin's law | | Weibull's law | |
|---|---|---|---|---|
| $\bar{b}$, MPa | 1,796.4 | b, MPa | 6,715 |
| $\bar{a}$ | -0.1 | a | -0.343 |

| $N_u$, cycles | 1,000 | B, cycles | 2,687 |
|---|---|---|---|
| $S_e$, MPa | 451.2 | $S_e$, MPa | 451.2 |

Table 1 – Example of material properties recalculation for a generic steel

The truncated Basquin's law and Weibull's law are shown in Figure 2; in the low cycle fatigue regime Weibull's law is more conservative than the truncated Basquin's law, while almost a decade after the inflection point, the smoothed curve becomes less conservative w.r.t. the truncated power law since the transition to the fatigue limit appears too slow. A possible way to recalibrate these constants in order to get a more conservative curve and/or a better fit of experimental data shall be explained hereafter:

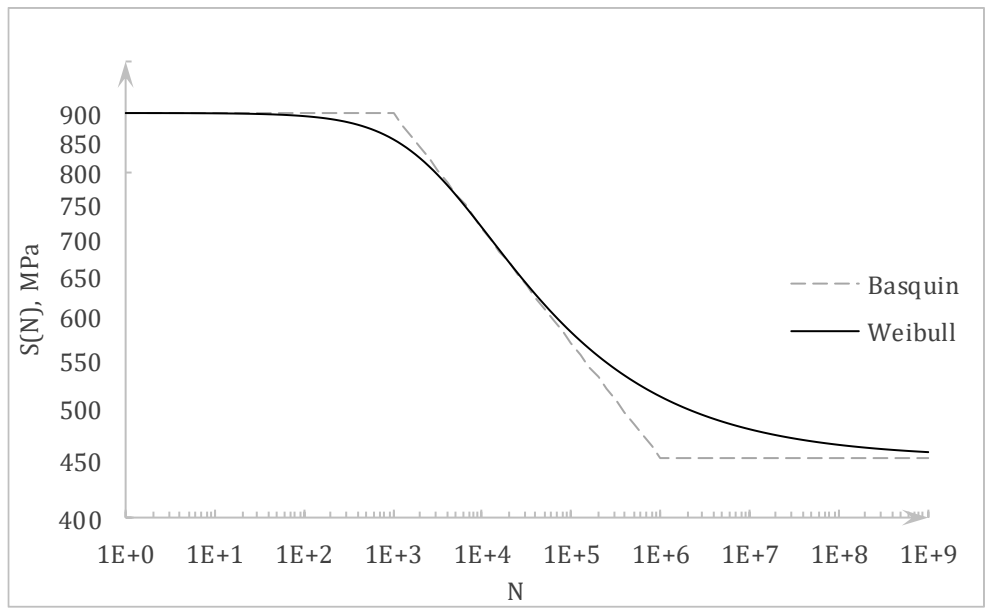

Figure 2 – Comparison between Basquin's law and the corresponding derived Weibull's law

### i. Techniques of recalibration of the constants

Graphical recalibration

If the experimental data require a steeper and more conservative curve in proximity of the fatigue limit, a smaller value of the exponent a should be chosen while keeping constant the inflection point $N_i$; on the contrary, if the transition to the fatigue limit is very smooth and should be modelled with a gentler slope, a higher value of the exponent should be used. In this way B, b, α vary consistently and the curve becomes more/less conservative in the very high cycle fatigue regime w.r.t. Basquin's truncated law, instead its first derivative for $N<N_i$ will not change too much. The definition of a slope factor $f_a$ accomplishes the scope of changing the slope in the high cycle fatigue regime, i.e.:

$$\begin{cases} ă = & f_a \cdot a & (a) \\ ă = & (a - ā)/(ā - 1) & (b) \\ B̆ = & N_i \cdot ă & (c) \\ b̆ = & (S_u - S_e) \cdot B̆^{-ă} & (d) \end{cases} \quad (15)$$

The family of curves deriving from changing the "slope" ă by multiplying a by 0<$f_a$<2 is shown in Figure 3.

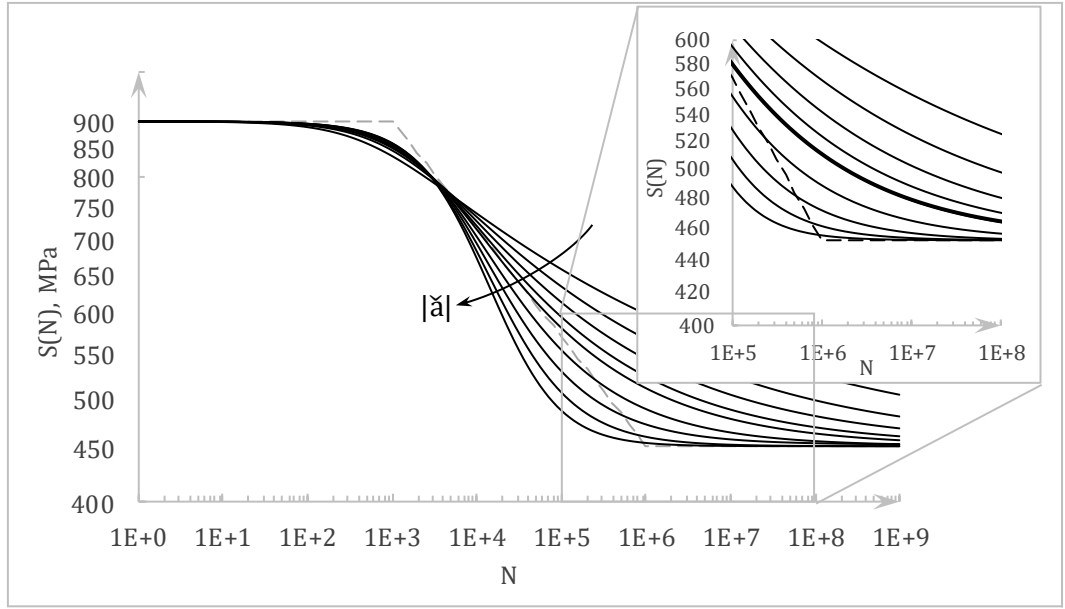

Figure 3 – Family of S/N curves deriving from the analytical fitting of the basquin's law with decreasing ă

## Statistical recalibration

The statistical calibration of the S/N curve parameters is usually done via least-squares method. The assumptions when approximating fatigue data with least squares by a pure power law are (i) that the N at prescribed S follows a lognormal distribution and (ii) that the variance of log(N) is constant over the tested range (hypothesis of homoscedasticity). The general equation of a lest squares regression line is

$$Y = C_1 X + C_2 + \epsilon \quad (16)$$

Where $\epsilon$ is a random variable of error. The regression line is then

$$\hat{Y} = \hat{C}_1 X + \hat{C}_2 \quad (17)$$

Where $\hat{C}_1, \hat{C}_2$ are the estimates obtained through the minimization of the sum of the squared deviations of the experimentally observed life from the predicted one considering $n_f$ tests

$$\Delta^2 = \sum_{j=1}^{n_f}\left(Y_j - \widehat{Y}_j\right)^2 = \sum_{j=1}^{n_f}\left(Y_j - (\widehat{C}_1 X_j + \widehat{C}_2)\right)^2 \tag{18}$$

From the minimization of $\Delta^2$ w.r.t. $\widehat{C}_1, \widehat{C}_2$, the estimated regression line is obtained:

$$\begin{cases} \widehat{C}_1 = \dfrac{\sum_{j=1}^{n_f}(X_j - \overline{X})(Y_j - \overline{Y})}{\sum_{j=1}^{n_f}(X_j - \overline{X})^2} & (a) \\ \widehat{C}_2 = \overline{Y} - \widehat{C}_1 \overline{X} & (b) \end{cases} \tag{19}$$

Where $\overline{X} = 1/n_f \sum_{j=1}^{n_f} X_j$, $\overline{Y} = 1/n_f \sum_{j=1}^{n_f} Y_j$ are the average values of X and Y. The variance of $Y_j$ on $X_j$ is

$$\sigma^2 = \frac{1}{n_f - 1} \sum_{j=1}^{n_f} \left(Y_j - (\widehat{C}_1 X_j + \widehat{C}_2)\right)^2 \tag{20}$$

Taking the decadic logarithm of the power law $S = \overline{b}\, N^{\overline{a}}$ gives $\text{Log}(N) = \frac{1}{\overline{a}}\text{Log}(S) + \frac{1}{\overline{a}}\text{Log}(b)$, hence Y=Log(N) and X=Log(S), from which Basquin's law constants are

$$\begin{cases} \overline{a} = 1/\widehat{C}_1 & (a) \\ \overline{b} = 10^{-\widehat{C}_2/\overline{a}} & (b) \end{cases} \tag{21}$$

From Lee et al. [39], the coefficient of variation, defined as the ratio of the standard deviation to the mean, is estimated as $C_{\overline{a}} = \sqrt{10^{\overline{a}^2 \cdot \sigma^2} - 1}$. The design S/N curve deriving from this approach is obtained with a confidence level (CL) of 50%. The simplest strategy to obtain a design S/N curve with higher CL is to introduce the lower limit $Y_L$ of Y at given X and level of confidence K$\sigma$, where K is a multiplier, i.e.

$$Y_{L_j}(K) = \widehat{Y}_j - K \cdot \sigma \tag{22}$$

Concerning the approximation of experimental data via Weibull's law and a required CL, the following three steps procedure could be applied: (i) calculate Basquin's law constants with the least squares method and the desired CL, (ii) calculate Weibull's law constants through equations (14) and (15) with $f_a$=1 and (iii) calibrate the slope factor $f_a$ through the least squares method considering Y=Log(N+B) and X=Log(S-$S_e$). In this way the smooth S/N curve can be represented as a power law from which $\breve{a} = 1/\widehat{C}_1$ and $\breve{b} = 10^{-\widehat{C}_2/\breve{a}}$ and the value $\breve{a}$ providing the 50% of CL can be found.

## ii. Notch effect

In the classical stress-life approach, the effect of notches can be accounted for through the definition of the material notch sensitivity factor q :

$$q = \frac{K_n - 1}{K_t - 1} \tag{23}$$

Where $K_n$ is the technical stress concentration factor, i.e. the predicted ratio of the plain endurance limit to that for the notched member, and $K_t$ is the theoretical elastic stress concentration factor. The notch sensitivity factor definition is based on Neuber's "building blocks" idea from 1946 [40], i.e. the material is not a continuum, but an aggregate of building blocks and the stress gradient cannot develop across blocks. Thus, a characteristic length $A_N$, equal to the half-size of the block, was defined such that the notch sensitivity factor takes the form

$$q = \frac{1}{1 + \sqrt{A_N/\rho}} \tag{24}$$

Where $\rho$ is the notch root radius and $A_N$ is Neuber's constant. In 1952 the effective stress concentration factor $K_f$ was experimentally measured for many materials by Kuhn and Hardrath [41] and also by Kuhn in a work from "Colloquium on Fatigue" [19] and good agreement was found with $K_n$. Thus, the approximation $K_n \approx K_f$ will be taken as valid. From Kuhn and Hardrath's work [41], the values of Neuber's constants giving the ratios $0.9 < K_n/K_f < 1.1$ are 0.02 in for aluminum alloys and 0.027 in for steels with $S_u \approx 115$ ksi. In 1949 Peterson [42] based on the approximation of a linear variation of the stress near the crack tip modified Neuber's equation (24) as

$$q = \frac{1}{1 + A_P/\rho} \tag{25}$$

Where $A_P$ is Peterson's material constant. Thus, by considering the notch effect, Basquin's law is then modified as also done by Nelson and Fuchs in "nominal stress range II method" [43]

$$\begin{cases} S(N) = & S_u = b_n \cdot N_u^{a_n} & 1 \leq N \leq N_u \\ S(N) = & b_n \cdot N_u^{a_n} & N_u < N \leq N_e \\ S(N) = & S_e/K_f = b_n \cdot N_u^{a_n} & N > N_e \end{cases} \tag{26}$$

Where $a_n = \bar{a} - \mathrm{Log}_{10}(K_f)/\mathrm{Log}_{10}(N_e/N_u)$ and $b_n = S_u/N_u^{a_n}$

Equation (26) can be used to obtain the same system of equations of (13) or (14) with $a_n$ and $b_n$ in place of $\bar{a}$ and $\bar{b}$. An example of application of Basquin's law smoothing for both plain and notched specimen is given in Figure 4.

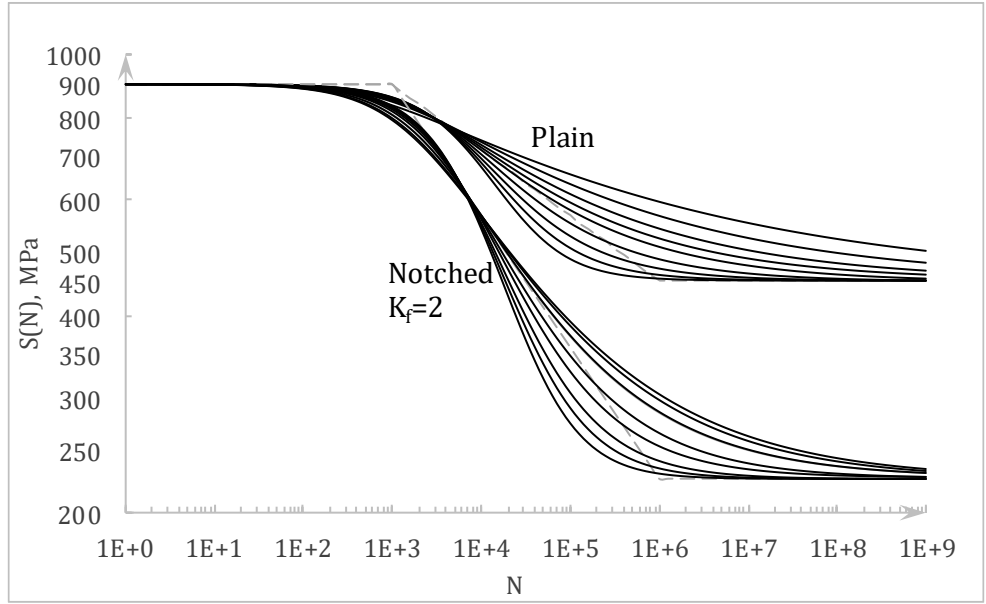

Figure 4 – Comparison between an S/N curve for plain and notched member

## 5. Quantitative validation with experimental data

In 1956 the National Advisory Committee for Aeronautics (NACA) conducted a fatigue test program on plain and notched 2024-T3, 7075-T6 aluminum alloy and SAE 4130 steel sheet specimens [44]. The fatigue data from this program were merged with other previous test programs [45]–[49] to gather the huge amount of S/N data collected in the NACA Technical Note 3866 [44]. The materials tensile properties are listed in Table 2. The note was aimed to fully characterize the S/N curve shape for all the materials tested, from $10^0$ to $10^8$ fatigue cycles and for $K_t$=1.0, $K_t$=2.0 and $K_t$=4.0 with emphasis on the range between 2 and 10,000 cycles. The geometry of the specimens under test is shown in Figure 5. The specimens were axially loaded and, depending on the stress amplitude, the loading frequency was opportunely adjusted from 12 (manual control) to 1,800 (subresonant testing) cycles per minute. In the last figure of the note Illg [44] showed a plot of $K_f$ vs the alternate stress observing that at very low number of cycles to failure $K_f \approx 1$ for all the materials testes, in good agreement with the S/N curve model later proposed by Nelson and Fuchs [43]. The maximum $K_f$ is found when the alternate stress tends to the fatigue limit; this value is generally in good agreement with Neuber's empirical equation (23). In this work the behavior of 2024-T3 aluminum alloy and normalized SAE 4130 steel have been considered in the case of fully reversed loading. Besides, results for 2024-T3 are applicable also to 7075-T6. In fact, *"due to the nearly identical fatigue properties of the T3, T351 and T4 conditions of 2024 and T6 and T651 conditions of 7075, no decision neds to be made between all these various conditions over the life region of interest"* [50].

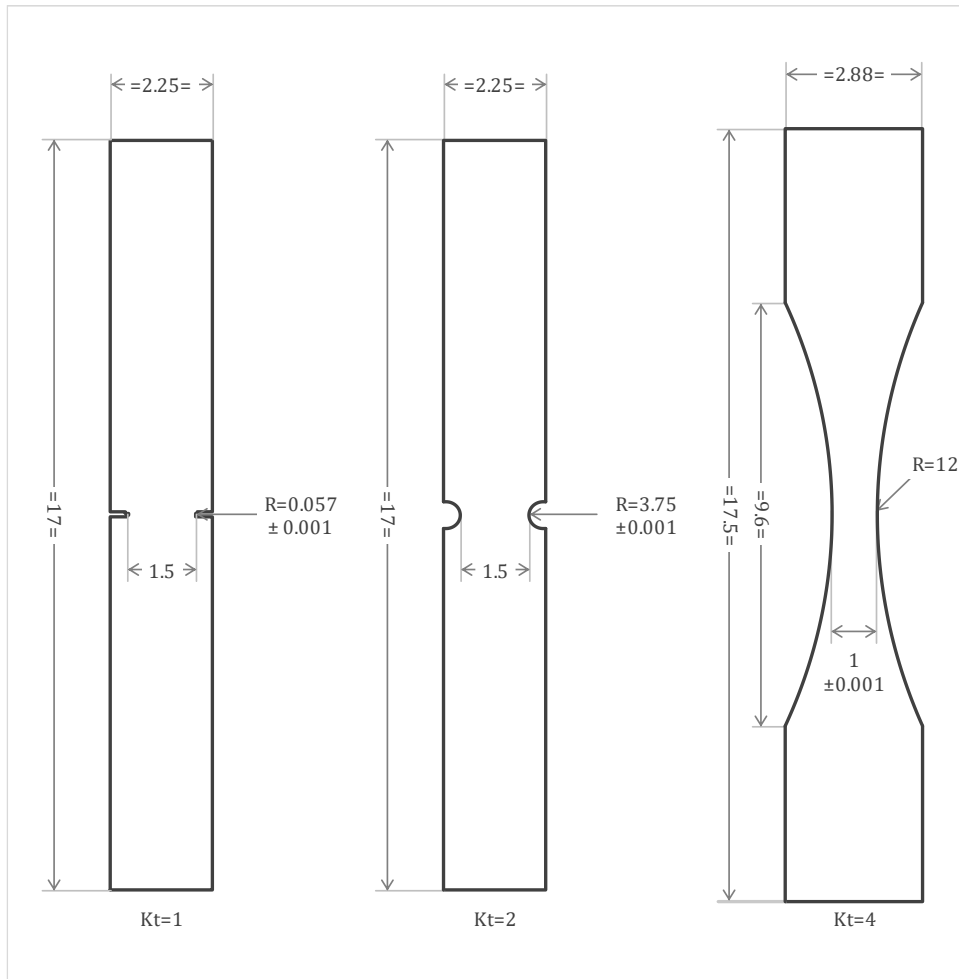

Figure 5– Configuration of sheet specimens, as in reference [44]. Lengths are in inches. Aluminum specimens are 0.09 inch thick; steel specimens are 0.075 inch thick

| Material | Number of tests | Yield stress, (0.2 percent offset), ksi | | | | Ultimate tensile strength, ksi | | | | Total elongation, 2-inch gage length, percent | | | | Young's modulus, ksi | | | |
|---|---|---|---|---|---|---|---|---|---|---|---|---|---|---|---|---|---|
| | | Av. | Min. | Max. | $\sigma^2$ (*) | Av. | Min. | Max. | $\sigma^2$ (*) | Av. | Min. | Max. | $\sigma^2$ (*) | Av. | Min. | Max. | $\sigma^2$ (*) |
| 2024-T3 aluminum alloy | 148 | 52 | 47 | 59.3 | 1.7 | 72.1 | 70.3 | 73.4 | 0.9 | 20 | 15 | 25 | 1.9 | 10,500 | 10,150 | 10,750 | 134 |
| 7075-T6 aluminum alloy | 152 | 76 | 71 | 79.8 | 1.4 | 83 | 79.8 | 84.5 | 1.1 | 12 | 7 | 15 | 1.3 | 10,200 | 10,000 | 10,550 | 104 |
| Normalized SAE 4130 steel | 149 | 94 | 87 | 102 | 2.1 | 116 | 111 | 125 | 1.8 | 15 | 12 | 18 | 1.1 | 29,400 | 28,200 | 31,500 | 660 |
| Hardened SAE 4130 steel | 9 | 174 | 168 | 178 | --- | 180 | 178 | 183 | --- | 8.3 | 8 | 9 | ---- | 29,900 | 29,200 | 30,800 | --- |

Table 2 – Tensile properties of the materials analyzed (from Illg's Table 1 [44]) (*) $\sigma^2$ is the standard deviation

Illg concludes his analysis with the following consideration: "The scatter in the results of the tests in the short-life range was remarkably small, whereas the tests at long lifetimes indicated considerably more scatter in the results" [44], in agreement with the capability of the model here described to vary consistently in proximity of the fatigue limit whilst keeping low scatter before the inflection point. All the S/N curves for the plain specimens have been calculated through the statistical procedure previously described with 50% of confidence level.

i. Plain specimens

As concerns the Al 2024-T3 plain specimens, from Table 2 $S_u$ = 72.1 ksi, while Basquin's law parameters have been calculated with a linear regression in log(S)/log(N) coordinates between $2 \cdot 10^3$ and $5 \cdot 10^6$ cycles, consistent with the trend observed from experimental data, giving $\bar{b}$ = 308.34 ksi and $\bar{a}$ = −0.191; consequently $S_e(N_e)$ = 16.32 ksi. Then, Weibull's law parameters have been calculated from equation (14), giving b = 1007.5 ksi, a = −0.347 and B = 4196. In order to obtain the Weibull's law with 50% of confidence level, equation (15) has been used to calculate and plot the linear regression in $Log(S-S_e)/Log(N+B)$ coordinates. The factor $f_a$ = 1.15 satisfies the required condition giving $\check{b}$ = 1743.9 ksi, $\check{a}$ = −0.399, and B = 5591.7. The S/N curve for normalized SAE 4130 steel has been calculated with the same procedure just introduced. All the fitting data are given in Table 3 and the final curves are shown in $Log(S-S_e)/Log(N+B)$ coordinates in Figure 6. For both the materials it is evident a dramatic increase in the scatter of fatigue data with decreasing stress amplitude, as also underlined by Illg [44]. Hence, in this case the hypothesis of homoscedasticity is maybe too rough to draw the Wöhler curves with the desired CL. Anyway, if the maximum variance is chosen to factorize the curve in stress (cfr. equation (22)), the estimate will certainly be conservative.

|  | Basquin |  | Weibull |  | Factorized Weibull |  |
|---|---|---|---|---|---|---|
| 2024-T3 aluminum alloy | $\bar{b}$, ksi | 308.3 | b, ksi | 1007.5 | $\check{b}$, ksi | 1744 |
|  | $\bar{a}$ | -0.191 | a | -0.3469 | $\check{a}$ | -0.3989 |
|  | $S_u$, ksi | 72.1 | B, cycles | 4196 | $\check{B}$, cycles | 5592 |
|  | $S_e$, ksi | 16.3 | $S_e$, ksi | 16.3 | $S_e$, ksi | 16.3 |
|  |  |  |  |  | Implies $f_a$ | 1.15 |
| Normalized SAE 4130 steel | $\bar{b}$, ksi | 161.2 | b, ksi | 159.393 | $\check{b}$, ksi | 266.3 |
|  | $\bar{a}$ | -0.0856 | a | -0.1773 | $\check{a}$ | -0.257 |
|  | $S_u$, ksi | 116 | B, cycles | 83 | $\check{B}$, cycles | 155 |
|  | $S_e$, ksi | 43 | $S_e$, ksi | 43 | $S_e$, ksi | 43 |
|  |  |  |  |  | Implies $f_a$ | 1.45 |

Table 3 – Fitting parameters for the S/N curve construction of the plain specimens

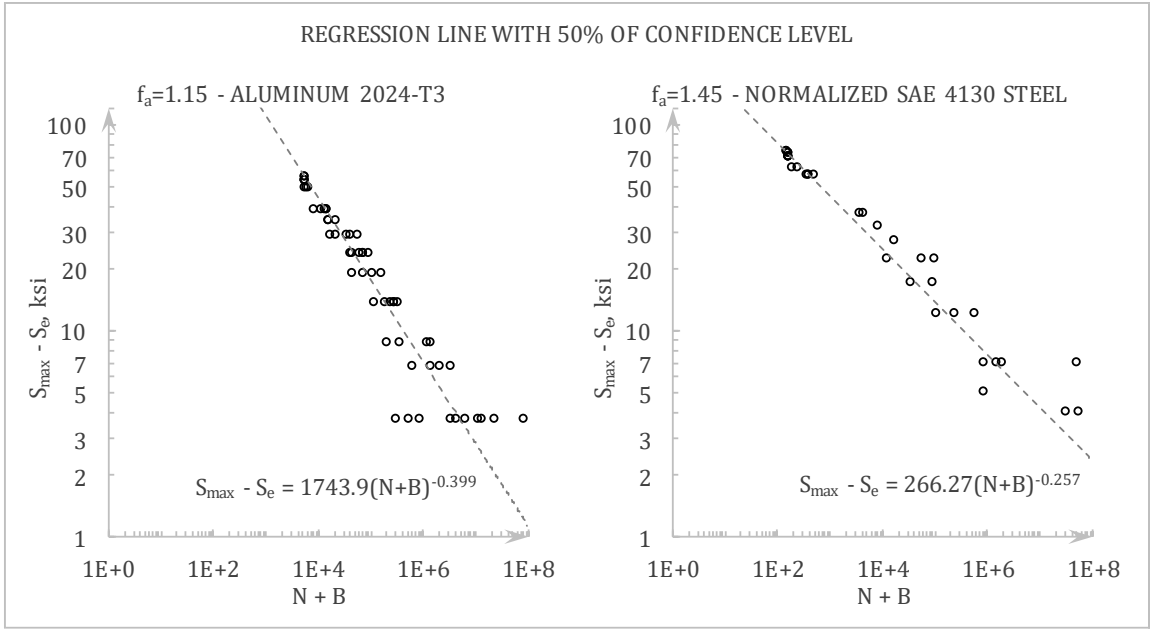

Figure 6 – Statistical recalibration of Weibull's law parameters via least squares method for plain specimens. The scaling $f_a$ used give 50% CL

## ii. Notched specimens

The S/N curves for the notched specimens have been calculated through the nominal stress range II method, i.e. equation (26), applied to the best fit truncated Basquin's law for the plain specimen. The values of $K_f$ have been calculated through equation (23) and then compared with Figure 16 of the NACA Technical Note 3866 [44] where $K_f$ is plotted vs the maximum stress. The values have been calculated through the Neuber's constants suggested by Kuhn and Hardrath's [41] which led to $K_f(K_t=2)\approx1.85$, $K_f(K_t=4)\approx2.90$ for aluminum and $K_f(K_t=2)\approx1.78$, $K_f(K_t=4)\approx2.78$ for steel. The effective stress concentration calculated for aluminum are confirmed by Figure 16 of the NACA Technical Note 3866 [44] and are also very similar to the values calculated by Topper et al. [50], while the values for steel at $K_t=4$ underestimated the actual notch effect, thence it had to be modified to $K_f(K_t=4)\approx3.65$ in accordance with Illg's Figure 16 [44]. Moreover, from Figure 16 [44] it is evident that the $K_f$ tends to one when the maximum nominal stress approaches the ultimate tensile strength, confirming the hypothesis that the effect of the presence of notches only marginally affects the ultimate tensile strength. Anyway, as regards Al 2024-T3, notched specimens do not show the initial plateau characterizing the curve for $K_t=1$, consequently the nominal stress range II approach as is does not fit properly the experimental data. For this reason, $N_u$ has been reduced from $2\cdot10^3$ to 20 for $K_t=2$ and to 2 for $K_t=4$, then the resulting Weibull's law has been recalibrated via equation (15). For normalized 4130 steel the nominal stress range II approach provided a satisfying fit of experimental data. The final S/N curves obtained through the model here introduced provide a smooth fit of the experimental data for both plain and notched experiments. The scaling factors remain unchanged with respect to the plain specimen, i.e. $f_a =$

1.15 and $f_a = 1.45$ for aluminum alloy and steel respectively. In Figure 7 the S/N curves obtained with the presented model are shown. Those curves are also compared with the non-corrected Weibull's law and with the truncated Basquin's law curves. The trend theorized by Weibull is globally confirmed, indeed Basquin's law through the present model appears to be an approximation of Weibull's law and the transition to the fatigue limit is smooth rather than a knee, especially when many data are available at $N \geq 10^7$ cycles.

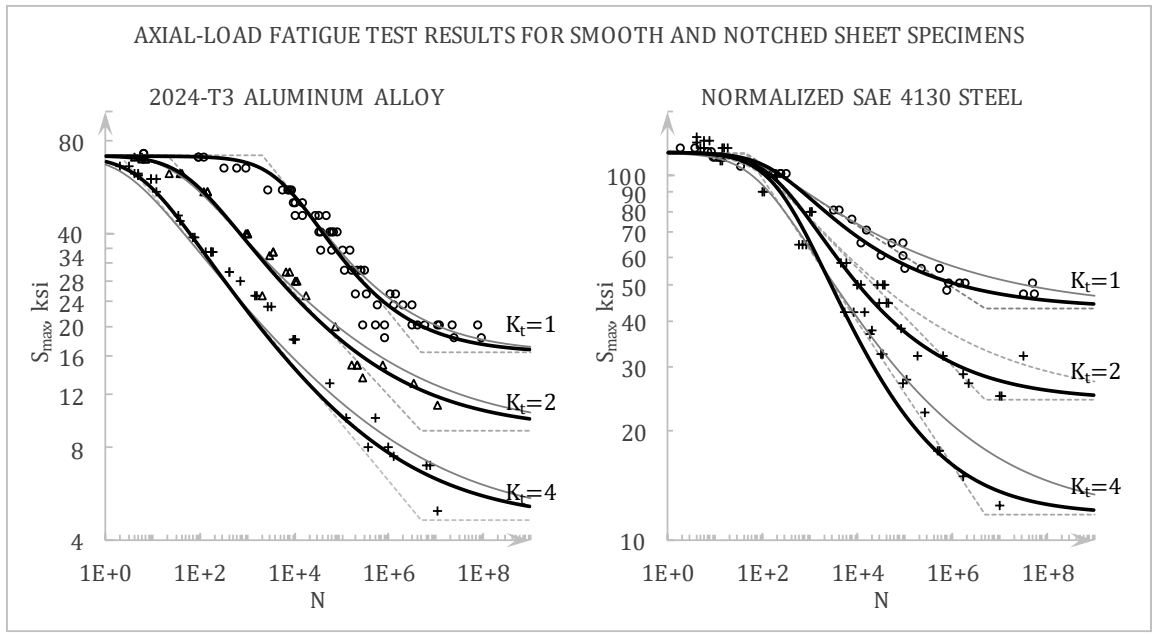

Figure 7 – S/N curves for plain and notched sheet specimens. The dashed gray line represents the truncated Basquin's law, the solid gray line is Weibull's law with no modifications, and the black solid line is the updated Weibull's law

## 6. Conclusions

An analytical explicit model relating Basquin's law truncated at the fatigue limit and Weibull's law for a smooth four parameters S/N curve is proposed. Weibull's approach for the "determination of average load-life relations" [31] made use of graphical or approximate analytical trial and error procedures to determine the S/N curve equation in semilogarithmic axes. The Log(S)/Log(N) coordinates are helpful since the Basquin's power law can be used to represent the slope in the inflection point, thus making straightforward the analytical exact definition of Weibull's law parameters. The model can be used to define a family of S/N curves where the scatter increases with decreasing stress amplitude through the only introduction of a slope factor $f_a$ and the hypothesis that $N_i$ stays constant. This result is in agreement with Illg's experiments [44] and shall be further studied from the statistical point of view. Finally, the effect of notches can be easily and successfully modelled only by knowing the effective stress intensity factor and the S/N curve for the plain specimen and it has been found that the parameter $f_a$ is not affected by the presence of the notch.

# Appendix

In order to define an analytical relationship between $S_U$, $\bar{a}$, $\bar{b}$ and a, b, B, the following definitions valid for the generic variable $\square$ have been used:

$$\log_{10} \square = \frac{\ln \square}{\ln 10} \tag{27}$$

Where ln stands for the natural logarithm.

Generic derivative of a quantity $\square$ with respect to the decadic logarithm of N

$$\frac{d\square}{d\log_{10} N} = \log 10 \frac{\frac{d\square}{dN}}{\frac{d \ln N}{dN}} = \ln 10 \cdot N \cdot \frac{d\square}{dN} \tag{28}$$

First derivative of the decadic logarithm of a function $\square(N)$ with respect to the decadic logarithm of N:

$$\frac{d \log_{10} \square(N)}{d \log_{10} N} = \log_{10} \square_{/\log_{10} N} = N \cdot \frac{d \ln \square}{dN} = N \cdot \frac{\square'(N)}{\square(N)} \tag{29}$$

Second derivative:

$$\begin{aligned}
\frac{d^2 \log_{10} \square(N)}{d(\log_{10} N)^2} &= \log_{10} \square_{/\log_{10} N \, \log_{10} N} = N \cdot \ln 10 \cdot \frac{d}{dN} \cdot \left[N \cdot \frac{\square'(N)}{\square(N)}\right] \\
&= N \cdot \ln 10 \cdot \frac{1}{\square(N)} \cdot \left[\square'(N) \cdot \left(1 - N \cdot \frac{\square'(N)}{\square(N)}\right) + N \cdot \square''(N)\right] \\
&= N \cdot \ln 10 \cdot \frac{1}{\square(N)} \cdot \left[\square'(N) \cdot \left(1 - \log_{10} \square_{/\log_{10} N}\right) + N \cdot \square''(N)\right]
\end{aligned} \tag{30}$$

Hence, substituting equation (5) into the generic derivatives written in equations (29) and (30) the following expressions are found:

$$\log_{10} S_{/\log_{10} N} = N \cdot \frac{b \cdot a \cdot (N + B)^{a-1}}{b \cdot (N + B)^a + S_e} \tag{31}$$

$$\begin{aligned}
\log_{10} & S_{/\log_{10} N \, \log_{10} N} \\
&= \frac{N \cdot \log 10 \cdot b \cdot a \cdot (N + B)^{a-2}}{b \cdot (N + B)^a + S_e} \\
&\quad \cdot \left[(N + B) \cdot \left(1 - \log_{10} S_{/\log_{10} N}\right) + N \cdot (a - 1)\right]
\end{aligned} \tag{32}$$